\documentclass[sigconf]{acmart}
\AtBeginDocument{%
  }

\setcopyright{acmlicensed}
\copyrightyear{2025}
\acmYear{2025}
\setcopyright{cc}
\setcctype{by}
\acmConference[CIKM '25]{Proceedings of the 34th ACM International
Conference on Information and Knowledge Management}{November 10--14,
2025}{Seoul, Republic of Korea}
\acmBooktitle{Proceedings of the 34th ACM International Conference on Information and Knowledge Management (CIKM '25), November 10--14, 2025,
Seoul, Republic of Korea}\acmDOI{10.1145/3746252.3760818}
\acmISBN{979-8-4007-2040-6/2025/11}
\usepackage{booktabs}   
\usepackage{tabularx}   
\usepackage{graphicx}   
\usepackage{caption}    
\usepackage{subcaption} 
\usepackage{array}

\begin{document}

\title{Fact or Facsimile? Evaluating the Factual Robustness of Modern Retrievers}

\author{Haoyu Wu}
\orcid{0009-0002-3097-7069}
\authornote{Both authors contributed equally to this research.}
\affiliation{%
  \institution{Northwestern University}
  \city{Evanston}
  \state{IL}
  \country{USA}
}
\email{haoyuwu2025@u.northwestern.edu}

\author{Qingcheng Zeng}
\orcid{0000-0002-8697-2729}
\authornotemark[1]
\affiliation{%
  \institution{Northwestern University}
  \city{Evanston}
  \state{IL}
  \country{USA}
}
\email{qingchengzeng2027@u.northwestern.edu}

\author{Kaize Ding}
\authornote{Corresponding Author.}
\affiliation{%
  \department{Department of Statistics and Data Science}
  \institution{Northwestern University}
  \city{Evanston}
  \state{IL}
  \country{USA}
}
\email{kaize.ding@northwestern.edu}






\renewcommand{\shortauthors}{Trovato et al.}

\begin{abstract}
Dense retrievers and rerankers are central to retrieval-augmented generation (RAG) pipelines, where accurately retrieving factual information is crucial for maintaining system trustworthiness and defending against RAG poisoning. However, little is known about how much factual competence these components inherit or lose from the large language models (LLMs) they are based on. We pair 12 publicly released embedding checkpoints with their original base LLMs and evaluate both sets on a factuality benchmark. Across every model evaluated, the embedding variants achieve markedly lower accuracy than their bases, with absolute drops ranging from 12 to 43 percentage points (median 28 pts) and typical retriever accuracies collapsing into the 25–35 \% band versus the 60–70 \% attained by the generative models. This degradation intensifies under a more demanding condition: when the candidate pool per question is expanded from four options to one thousand, the strongest retriever’s top-1 accuracy falls from 33 \% to 26 \%, revealing acute sensitivity to distractor volume. Statistical tests further show that, for every embedding model, cosine-similarity scores between queries and \emph{correct} completions are significantly higher than those for \emph{incorrect} ones ($p < 0.01$), indicating decisions driven largely by surface-level semantic proximity rather than factual reasoning. To probe this weakness, we employed GPT-4.1 to paraphrase each correct completion, creating a rewritten test set that preserved factual truth while masking lexical cues, and observed that over two-thirds of previously correct predictions flipped to wrong, reducing overall accuracy to roughly one-third of its original level. Taken together, these findings reveal a systematic trade-off introduced by contrastive learning for retrievers: gains in semantic retrieval are paid for with losses in parametric factual knowledge, and the resulting models remain highly vulnerable to adversarial or even benign rephrasings. Our study underscores the need for retrieval objectives that balance similarity with factual fidelity to safeguard next-generation RAG systems against both misinformation and targeted attacks.

\end{abstract}



\maketitle

\section{Introduction}

RAG has become a go-to strategy for grounding LLMs in external evidence, sharply reducing hallucination rates across question answering, dialogue, and long-form generation. Early work showed that coupling a seq-to-seq generator with a neural retriever yields more specific and truthful answers than parametric models alone\,\citep{lewis2021retrievalaugmentedgenerationknowledgeintensivenlp,shuster-etal-2021-retrieval-augmentation, borgeaud2022improvinglanguagemodelsretrieving}. A contemporary RAG pipeline contains three cooperative modules: \textbf{(1) a retriever} that selects top-$k$ passages from a database, typically via dense dual-encoder models such as Dense Passage Retrieval (DPR)\cite{karpukhin-etal-2020-dense}; \textbf{(2) an optional reranker}---often a cross-encoder based on LLMs---that refines this shortlist for precision; and \textbf{(3) an LLM generator} that conditions on the evidence to produce the final answer. This two-stage retrieval stack trades speed in the first step for accuracy in the second, delivering state-of-the-art recall without sacrificing fluency\,\citep{nogueira2020passagererankingbert}.

Modern dense retrievers are trained with \emph{contrastive objectives} that pull queries toward semantically similar passages while pushing dissimilar ones apart. For instance, the DPR model trains dual BERT encoders for questions and passages such that the embedding of a question has a high inner-product similarity with its true answer passage and lower similarity with unrelated passages \cite{karpukhin2020densepassageretrievalopendomain}. This contrastive approach effectively narrows the semantic gap between queries and corresponding documents, enabling the retriever to find relevant information even when the query uses different wording than the documents. However, optimizing for semantic similarity can also introduce an over-reliance on surface matching. The retriever may pick documents that \emph{sound} relevant to the query---sharing many keywords or general topic overlap---but which do not actually contain the correct answer or factual details needed \cite{sriram2024contrastivelearningimproveretrieval}. Recent analyses document systematic ``short, early, and literal'' biases that undermine factual recall when the surface form and truth diverge\,\citep{karpukhin-etal-2020-dense, fayyaz2025collapsedenseretrieversshort}.

Crucially, most RAG systems assume that whatever is retrieved is both relevant \emph{and} reliable. Emerging security work shows this assumption is brittle: injecting only a handful of poisoned passages can hijack the retriever and steer otherwise well-behaved LLMs toward false or harmful content\,\citep{xue2024badragidentifyingvulnerabilitiesretrieval}. Yet empirical evidence on how such retrieval noise affects factual accuracy---especially relative to the base LLMs without retrieval---remains sparse. In this study, we explicitly measure a suite of dense retrievers and rerankers (hereafter IR models) and their corresponding base LLMs on the same factuality benchmark, and our results suggest a uniform \textit{factuality degradation} on all embedding models. Further analysis even reveals that for correctly answered questions, IR models rely more on similarity rather than factuality reasoning. To sum up, our contribution could be summarized as follows:
\begin{enumerate}
    \item \textbf{Quantifying factuality drift.} We explicitly measure accuracy gaps between several widely-used IR models checkpoints and their corresponding base models, showing a consistent factuality decrease.
    \item \textbf{Providing a rigorous statistical analysis.} We couple aggregate accuracy drops with permutation tests and bootstrap confidence intervals to confirm that the observed degradation is both significant and robust across model families.
    \item \textbf{Revealing a potential RAG vulnerability.} Our findings expose an overlooked trade-off in today’s RAG pipelines---diminished factual fidelity in retrieval components both enlarges the attack surface and underscores the need for \emph{factuality-aware retrieval training}, as adversaries can slip in near-identical distractors that the IR models favors, causing the generator to hallucinate with unwarranted confidence.

\end{enumerate}

\section{Models and Dataset}
To systematically quantify the degree to which factual knowledge is retained when an LLM undergoes fine-tuning as a retriever or reranker, we pair each IR model with the corresponding \emph{base LLM} from which it was originally derived. This pairing allows us to directly compare the factuality capabilities of the downstream model with its base version, isolating the impact of the fine-tuning process on factual knowledge. For model selection, we focus on top-performing retrievers listed on the MTEB leaderboard \cite{muennighoff2023mtebmassivetextembedding}, which provides a comprehensive evaluation of embedding models across various retrieval benchmarks. For each selected IR models, we identify and include its associated base model in our evaluation. A detailed summary of all evaluated models is provided in \autoref{tab:model}.

\begin{table}[ht]
  \centering
  \small{
  \begin{tabularx}{\linewidth}{@{}l l X@{}}
    \toprule
    \textbf{Base LLM} & \textbf{Version} & \textbf{Retriever/Reranker} \\
    \midrule
    Mistral-7B-v0.1 & 7 B &
      \begin{tabular}[l]{@{}l@{}}
        e5-mistral-7b-instruct\\
        Linq-Embed-Mistral\\
        SFR-Embedding-2\_R\\
        SFR-Embedding-Mistral
      \end{tabular} \\[4pt] \hline
    Qwen2-7B & 7 B &
      gte-Qwen2-7B-instruct \\[4pt] \hline
    Qwen2-1.5B & 1.5 B &
      \begin{tabular}[l]{@{}l@{}}
        gte-Qwen2-1.5B-instruct\\
        stella\_en\_1.5B\_v5
      \end{tabular} \\[4pt] \hline
    Qwen2-0.5B & 0.5 B &
      KaLM-embedding-multilingual-mini-v1 \\[4pt] \hline
    Mistral-7B-v0.2 & 7 B &
      FollowIR-7B (reranker) \\[4pt] \hline
    gemma-2-9B & 9 B &
      bge-reranker-v2.5-gemma2-lig \\[4pt] \hline
    Qwen2.5-1.5B & 1.5 B &
      mxbai-rerank-large-v2 \\[4pt] \hline
    LLama-3.1-8B & 8 B & 
    ReasonIR-8B \\[4pt] \hline
    \bottomrule
  \end{tabularx}
  }
  \caption{Base LLMs and their corresponding retriever or reranker models}
  \label{tab:model}
\end{table}

Our evaluation is conducted on the \textbf{FACTOR} benchmark \cite{muhlgay2024generatingbenchmarksfactualityevaluation}, which is specifically designed to probe the factual accuracy of embedding models. Each instance in FACTOR is structured as a four-way multiple-choice question comprising the following elements:

\begin{itemize}
\item \textbf{Prefix}: a concise factual prompt (for example, “The capital of Canada is \ldots”).
\item \textbf{Completions (A–D)}: one factually correct sentence and three minimally modified distractors. Each distractor is crafted to include a single, clearly defined error type: \textit{Entity}, \textit{Predicate}, \textit{Circumstance}, \textit{Coreference}, or \textit{Link}. This controlled construction makes the task highly challenging for embedding models, as distinctions between choices often hinge on subtle changes at the token level.
\end{itemize}

\begin{table}[htbp]
    \centering
    \small{
    \begin{tabularx}{\linewidth}{@{} l X @{}}
        \toprule
        \textbf{Item} & \textbf{String} \\ \midrule
        Prefix & Telegraph Act is a stock short title which used to be used for legislation in the United Kingdom, relating to telegraphy. \\ \hline
        Correct Completions & The Bill for an Act with this short title may have been known as a Telegraph Bill during its passage through Parliament. \\ \hline
        Contradiction & The Bill for an Act with this short title may have been known as a Wireless Bill during its passage through Parliament. \\ \hline
        Rewritten Correct Completion & During its progression in Parliament, legislation bearing this short title might have been referred to as the Telegraph Bill. \\
        \bottomrule
    \end{tabularx}}
    \caption{Sample case}
    \label{tab:prefix-completions}
\end{table}

\begin{figure*}[htbp]          
  \centering
  \includegraphics[width=\textwidth]{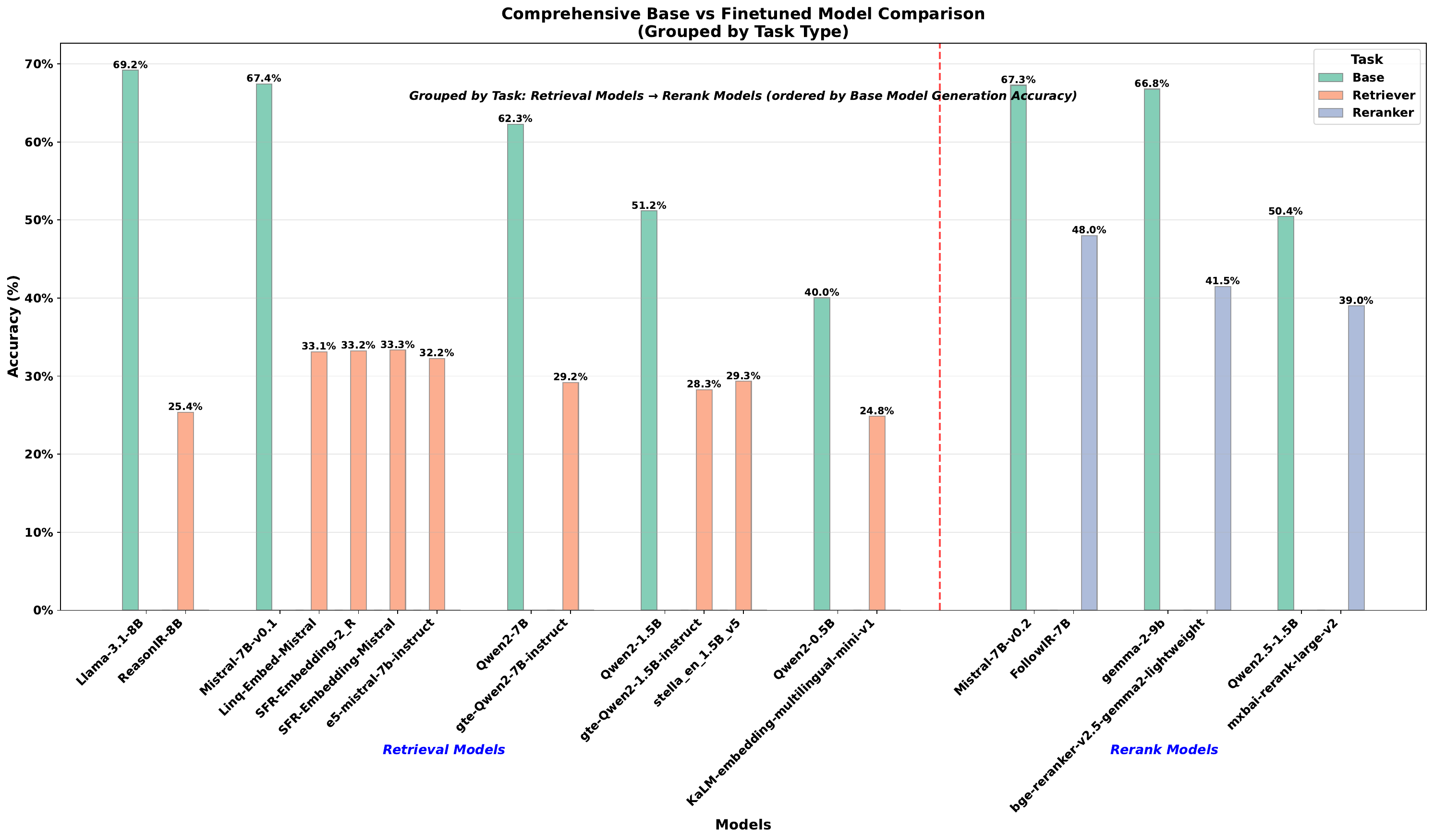}%
  \vspace{-6pt}
  \caption{comprehensive base model comparison}      
  \label{fig:model_comparision}
\end{figure*}

\section{Results}
\subsection{General Evaluation}

\subsubsection{Evaluation Setup}
For base LLMs, we follow the evaluation protocol established by the original FACTOR benchmark. Specifically, for each of the four candidate completions, we concatenate it to the given prefix and compute the perplexity of the resulting sequence using the base LLM. Formally, for a candidate sequence $w$, the perplexity is given by:
\begin{equation}
\mathrm{Perplexity}(P) = \exp\left(-\frac{1}{N} \sum_{i=1}^{N} \log P(w_i) \right)
\end{equation}
where $N$ denotes the total number of tokens in the sequence, and $P(w_i)$ represents the probability assigned to the 
$i$-th token by the LLM. For each question, we select the completion that yields the lowest perplexity as the model’s prediction, reflecting the choice that the base LLM deems most probable and thus, presumably, most factual.

For IR models, we adapt the evaluation approach to match their architecture. The prefix is used as the query, preceded by an instruction: \textit{From the following completions, find the factually correct one for the following prefix}. For retrievers, we independently embed the query and each candidate completion, then select the completion whose embedding has the highest cosine similarity with the query embedding as the final answer. For rerankers, we compute a relevance score between the query and each candidate completion using the reranker model, and select the completion with the highest relevance score as the prediction. We use accuracy as our primary evaluation metric. For each question, the model’s selected completion is considered correct if it matches the ground-truth answer provided by the FACTOR benchmark. The overall accuracy is then calculated as the proportion of correctly answered questions across the entire dataset, providing a straightforward measure of the factuality performance of each evaluated model.
\subsubsection{Results}
Our comparative analysis, as illustrated in the figure \ref{fig:model_comparision}, reveals a consistent and significant performance disparity between base generative models and their specialized counterparts for retrieval and reranking. The core finding is a notable degradation in factuality assessment capabilities when a base model is fine-tuned for a specific downstream task.

Across diverse model families such as Llama, Mistral, and Qwen, the base models demonstrate high accuracy (typically 60-70\%) in identifying factually correct information from a set of choices. In stark contrast, their corresponding retriever models, which are fine-tuned for dense retrieval, exhibit a sharp drop in performance, with accuracy often falling to the 25-35\% range. This trend strongly suggests that the fine-tuning process for semantic retrieval may inadvertently erode the rich factual knowledge encapsulated within the original base models.

A similar pattern of performance degradation is observed for reranker models. While models like bge-reranker-v2.5-gemma2-lightweight and mxbai-rerank-large-v2 are designed to refine search results, they still underperform their respective base models (e.g., gemma-2-9b and Mistral-7B-v0.2) on this factuality benchmark. The accuracy of rerankers, while in some cases higher than retrievers, remains substantially lower than that of the base models.

\subsection{Retriever-setting Evaluation}
\paragraph{Setup}We selected the SFR-Embedding-Mistral retriever because it achieved the highest Top-1 accuracy among all retrievers in the Figure \ref{fig:model_comparision}. To evaluate it in a setting that more closely mirrors conventional, large-scale applications, we modified the standard evaluation framework. We expanded the candidate pool from 4 options to 1000, where the 996 additional "distractor" candidates were randomly sampled from completions of other questions in the dataset. This high-cardinality setup provides a more realistic test of the model's ability to identify the single correct answer from a vast number of plausible alternatives.
\paragraph{Results} Under this expanded candidate pool, the performance of SFR-Embedding-Mistral degraded significantly. Its Top-1 accuracy dropped from 33.3\% in the standard 4-option task to just 26.6\%. This result is also substantially lower than the 67.4\% accuracy achieved by the baseline model. The retriever's significant performance degradation underscores the widespread challenge of factuality degradation, where a model’s ability to discern the correct answer diminishes as the candidate space grows—a problem not inherent to perplexity-based methods.

\section{Analysis and Discussion}
In the previous section, we demonstrated that LLM-based retrievers are substantially less effective than their original base models at performing factuality reasoning. In this section, we further investigate the mechanisms underlying IR models’ performance on factuality tasks. Specifically, we explore whether IR models answer factuality questions primarily by leveraging surface-level similarity or by engaging in actual factuality reasoning.

\subsection{Statistical Analysis}
Ideally, if IR models are genuinely performing factuality-based reasoning, the similarity between the query and the correct completions should be comparable across both sets of correctly and incorrectly answered questions. In other words, the IR models’s ability to answer factuality questions should not be solely determined by superficial similarity, but by deeper reasoning about factual content. To explore this, we compare both sets of similarities and use the Mann–Whitney U test \cite{mann1947test} to test the statistical significance.

\begin{table}[h]
  \centering
  \tiny{
  \begin{tabular}{@{}l l l | c | l l l@{}}
    \toprule
    \textbf{Model} & \textbf{Correct} & \textbf{Incorrect} & & \textbf{Model} & \textbf{Correct} & \textbf{Incorrect} \\
    \midrule
    SFR-Embedding-Mistral    & 0.678** &  0.652 & & FollowIR-7B    & 0.999** &  0.998 \\
    mxbai-rerank-large-v2   & 8.279**   &  7.889 & & KaLM-embedding   & 0.521** & 0.487 \\
    bge-reranker-v2.5-gemma2 & 20.794** & 20.159 & & gte-Qwen2-1.5B-instruct  & 0.483** &  0.439 \\
    ReasonIR-8B   & 0.479**   &  0.453 & & gte-Qwen2-7B-instruct  & 0.467** &  0.436 \\
    \bottomrule
  \end{tabular}}
  \caption{The comparison of similarities between correctly and incorrectly answered questions. ** indicate $p<0.01$.}
  \label{tab:analysis}
\end{table}

Our results for each IR model, as presented in \autoref{tab:analysis}, indicate that a consistent and statistically significant pattern (p < 0.01) across all evaluated IR models: the similarity scores for ground-truth documents are significantly higher when they are correctly ranked. This indicates a strong positive correlation between a model's assigned relevance score and its predictive accuracy. Consequently, this pattern provides partial evidence that IR models may not be relying on genuine factuality reasoning, but instead are making decisions based on superficial similarities.

\subsection{Paraphrase Attack}
\begin{figure}[htbp]
  \centering
  \includegraphics[width=\linewidth]{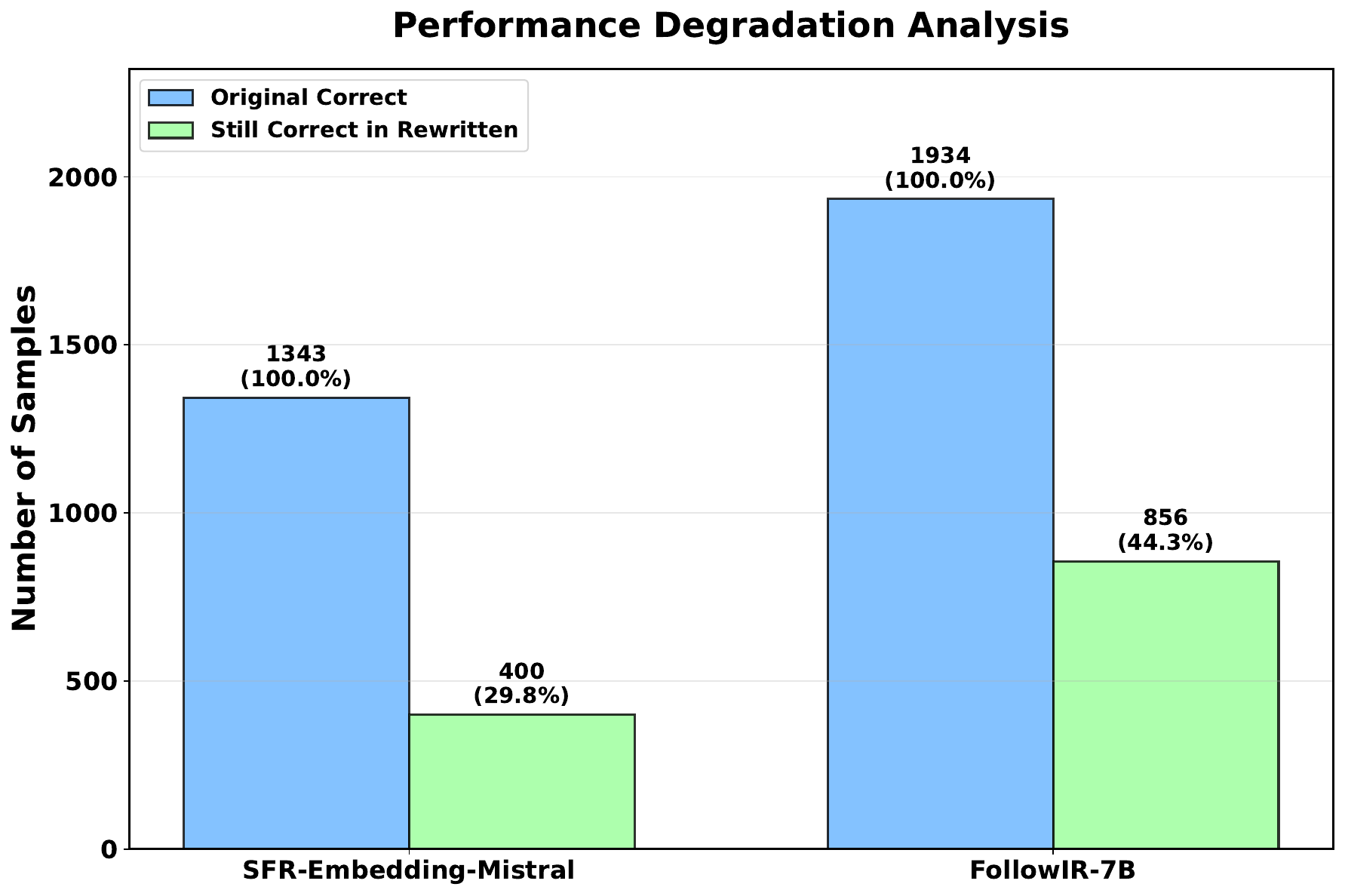}%
  \vspace{-6pt}
  \caption{performance degradation}      
  \label{fig:performance_degradation}
\end{figure}

To better estimate the causal effect of similarity, we employ a paraphrasing-based attack on the model. Specifically, we prompt GPT-4.1 \cite{openai2024gpt4ocard} to paraphrase the correct completion so that it expresses the same meaning as the original answer but is as dissimilar to the query as possible. This process minimizes the similarity signal while preserving factual content. To assess the effectiveness of this approach, two authors conducted a case study on 50 instances and confirmed that GPT-4.1 could reliably generate such paraphrases, with the reduced similarity reflected by lower Jaccard similarity scores. One example rewritten completion is shown in \autoref{tab:prefix-completions}. Ideally, if IR models are capable of detecting factuality, they should still be able to correctly identify the paraphrased answers for questions they originally answered correctly. To validate this approach, we conduct a case study using one retriever (SFR-Embedding-Mistral) and one reranker (mxbai-rerank-large-v2) to examine how their performance changes on questions they previously answered correctly after the correct completions are paraphrased. Our results are shown in \autoref{fig:performance_degradation}.

\vspace{-6pt}
\vspace{-6pt}

These results provide more convincing evidence that IR models are explicitly optimized for semantic similarity rather than for factuality. For both models, accuracy drops to around 30\% on questions they originally answered correctly after paraphrasing. This suggests that current optimization strategies may fail to equip models with genuine factual reasoning ability, limiting their reliability in tasks that require robust factual understanding.

\subsection{Discussion}
In this paper, we investigate whether IR models are capable of factuality-based retrieval compared to their base counterparts. Our results show that mainstream IR models experience significant accuracy drops relative to the base models. Moreover, our analysis reveals that even for questions answered correctly, IR models primarily depend on semantic similarity rather than factuality-based reasoning.

Recently, a series of works \cite{ECIR_RAG_ATTACK, cho-etal-2024-typos, zhou2025trustragenhancingrobustnesstrustworthiness} have highlighted the importance of trustworthy RAG and demonstrated various ways in which RAG systems can be attacked. Our results offer a new perspective on the fragility of RAG, suggesting that IR models may overprioritize semantic similarity, which leads to weaker defenses against factuality-based attacks on the RAG database. These findings underscore the need to reconsider how IR models are optimized within RAG systems. Prioritizing semantic similarity alone is insufficient for ensuring factual reliability, as it exposes the system to simple paraphrasing attacks that can undermine retrieval accuracy. Strengthening RAG against such vulnerabilities will require new training objectives or architectural changes that explicitly incorporate factuality reasoning. Addressing this gap is crucial for deploying RAG systems in real-world applications where trustworthiness and factual correctness are essential.

\section*{GenAI Disclosure}
During the preparation of this work, the authors utilized generative AI tools, including OpenAI's ChatGPT and Cursor, to assist with language editing, grammar correction, and code implementation. The authors have meticulously reviewed and edited all AI-generated content and assume full responsibility for the final manuscript.

\bibliographystyle{ACM-Reference-Format}
\bibliography{reference}

\end{document}